\documentclass[aps,twocolumn,showpacs,showkeys,preprintnumbers,superscriptaddress,nobibnotes,floatfix,nofootinbib]{revtex4-2}

\usepackage{orcidlink}
\usepackage{amsmath}
\usepackage{amsfonts}
\usepackage{amssymb}
\usepackage{mathrsfs}
\usepackage{placeins}          % \FloatBarrier
\usepackage{bm}                % \bm{\alpha}
\usepackage{bbm}               % \mathbbm{1}
\usepackage{slashed}           % \slashed
\usepackage{physics}           % \grad, \curl
\usepackage{esvect}            % \vec{v}
\usepackage[version=4]{mhchem}

\usepackage{xcolor}
\usepackage{graphicx}
\usepackage{array}
\usepackage{url}
\usepackage{amsthm}
\usepackage{enumerate}
\usepackage{enumitem}
\usepackage{booktabs}          % \toprule, \midrule, \bottomrule
\usepackage{siunitx}
\usepackage{mdframed}
\usepackage{tcolorbox}
\usepackage{fontawesome5}
\usepackage{braket}
\usepackage{setspace}
\usepackage{multirow}
\usepackage{upgreek}
\usepackage{xspace}
\usepackage[normalem]{ulem}    % \sout

\usepackage{natbib}
\usepackage{hyperref}
\hypersetup{
    colorlinks=true,
    linkcolor=blue,
    citecolor=blue,
    filecolor=magenta,
    urlcolor=blue,
    breaklinks=true
}
\usepackage[all]{hypcap}
\usepackage[capitalise]{cleveref}
\usepackage{lineno}            % line numbers for review, see \linenumbers below

\Crefname{section}{Sec.}{Secs.}
\Crefname{table}{Tab.}{Tabs.}
\Crefname{figure}{Fig.}{Figs.}
\Crefname{equation}{Eq.}{Eqs.}

\setlist[description]{leftmargin=\parindent,labelindent=\parindent}

\begin{document}
%\linenumbers % For review only
\hfill TTP26-038

\title{Learning Astrophysical Uncertainties in Dark Matter Direct Detection with Simulation-Based Inference}%

\author{Felix Kahlhoefer}
\email{felix.kahlhoefer@kit.edu}
\affiliation{Institute for Astroparticle Physics (IAP), Karlsruhe Institute of Technology (KIT),
Hermann-von-Helmholtz-Platz 1, 76344 Eggenstein-Leopoldshafen, Germany}
\affiliation{Institute for Theoretical Particle Physics (TTP), Karlsruhe Institute of Technology (KIT), 76128 Karlsruhe, Germany}

\author{Niklas Reus}
\email{niklas.reus@kit.edu}
\affiliation{Institute for Astroparticle Physics (IAP), Karlsruhe Institute of Technology (KIT),
Hermann-von-Helmholtz-Platz 1, 76344 Eggenstein-Leopoldshafen, Germany}

\date{\today}

%% =====================================================================
\begin{abstract}
We present a proof-of-concept study of neural ratio estimation~(NRE)~--- a
simulation-based inference algorithm~--- applied to WIMP direct detection using
the XENONnT experiment as an example. A binary classifier is trained on
simulated parameter--observation pairs to approximate the likelihood ratio for
the WIMP mass~$m_\chi$ and spin-independent proton coupling~$c_p$, using
realistic scintillation and ionization signals produced by the \texttt{fuse}
detector simulation. We demonstrate that astrophysical uncertainties from halo
model variations are marginalized over implicitly and at no additional inference
cost by including multiple halo models in the training data. Using extreme
$\pm5\sigma$ Standard Halo Model variations, we show that training on a combined
dataset recovers well-calibrated posteriors even under severe misspecification,
while more realistic halo models produce nearly indistinguishable results. The
signal-plus-background classifier yields an exclusion limit that is both
qualitatively and (after accounting for differences between frequentist and
Bayesian approaches) quantitatively consistent with the XENONnT expected
sensitivity. These results establish NRE as a scalable complement to the profile
likelihood ratio approach, with particular advantages as experiments probe
deeper into the neutrino fog.
\end{abstract}

\maketitle

%% =====================================================================
\section{Introduction}
\label{sec:intro}
%% =====================================================================

Understanding the particle nature of dark matter~(DM) is among the central
challenges in modern physics. Within the $\Lambda$CDM cosmological framework, DM
accounts for approximately 27\% of the total energy budget of the present
universe~\cite{PLANCK_2020_Planck-2018-Results}, yet no particle within the
Standard Model~(SM) can explain its observed properties. Weakly Interacting
Massive Particles~(WIMPs) remain among the most theoretically compelling
candidates~\cite{Feng_2023_WIMPs}: a thermally produced particle with weak-scale
mass and couplings naturally reproduces the observed relic abundance via
freeze-out.

Direct detection experiments search for the rare scattering of galactic WIMPs
off nuclei, converting the resulting nuclear recoil energy into measurable
signals. Leading experiments based on dual-phase liquid xenon~(LXe) time
projection chambers~--- XENONnT~\cite{XENON_2024_The-XENONnT-DM-Experiment},
LZ~\cite{LZ_2019_The-LZ-experiment}, and
PandaX-4T~\cite{PandaX_2024_PandaX4T}~--- exploit the simultaneous detection of
scintillation~(S1) and ionization~(S2) signals for three-dimensional position
reconstruction and powerful discrimination between nuclear recoil~(NR) signal
and electron recoil~(ER) background. Despite decades of experimental effort, no
convincing WIMP signal has been observed. The most stringent current bounds on
the spin-independent~(SI) WIMP--nucleon cross section reach
$\sigma_p = 2.2\times10^{-48}~\mathrm{cm}^2$ at
$40~\mathrm{GeV}/c^2$~\cite{LZ_2025_Results}.

The conventional statistical framework for WIMP searches is the extended profile
likelihood ratio~(PLR) test, in which signal and background probability density
functions are built from detector simulations and calibration data, and nuisance
parameters encoding systematic uncertainties are profiled over. To reduce
complexity, astrophysical uncertainties from the local DM velocity distribution
are commonly not included and the relevant parameters are simply fixed to the
Standard Halo Model~(SHM) central values without profiling. As the sensitivities
of direct detection experiments improve this approximation will need to be
revisited~\cite{Baxter_2021_DMDD-Recommendations}. However, each additional
nuisance parameter increases the dimensionality of the profiling problem and the
computational cost of toy-Monte Carlo~(toyMC) based inference, which already
requires $10^4$--$10^5$ pseudo-experiments per tested cross section to
characterize the sampling distribution of the test statistic.

Simulation-based inference~(SBI)~\cite{Cranmer_2020_SBI,
Deistler_2025_SBI-Guide}, originally named likelihood-free inference, addresses
this challenge by training neural networks on simulated parameter--observation
pairs~$(\boldsymbol{\vartheta}, \mathbf{x})$ to learn a surrogate for the
likelihood ratio or posterior directly, without requiring an explicit expression
for the likelihood. Once trained, inference is \emph{amortized}: the network
evaluates new observations at negligible cost. A particularly attractive feature
is that nuisance parameters are absorbed implicitly into the training
distribution by sampling them from their allowed ranges at simulation time and
discarding them before training. The resulting marginalization is exact in the
infinite-data limit and requires no modification of the inference
step~\cite{Cranmer_2020_SBI}. SBI has increasingly been used in searches for
dark matter~\cite{Cerdeno:2024uqt,Decant:2024bpg, Nguyen_2026_SBI_Dwarf, Montel_2022_NRE_WDM,
Hermans_2021_WDM_NSBI, Simola_2019_ML_DMDD,Aalbers:2024qqb}, but also in other
branches of physics, from gravitational wave astronomy~\cite{Liang_2025_SBI-GWs}
to cosmological parameter inference~\cite{Lemos_2023_SBI-Cosmo} and searches for
physics beyond the Standard
Model~\cite{Chathirathas_2024_Findingexcessesmodelparameter,
Morandini_2024_Axions}.

In this paper, we present a proof-of-concept application of neural ratio
estimation~(NRE), a specific SBI algorithm that approximates the likelihood
ratio via a binary classifier, using the XENONnT experiment as an example.\footnote{A similar study was performed in Ref.~\cite{Cerdeno:2024uqt} with a focus on parameter reconstruction from future data sets. Our goal instead is to obtain robust exclusion limits from current data sets.} We
use a realistic signal model with WIMP spectra generated by
WimPyDD~\cite{Jeong_2022_WimPyDD} and detector response modeled by the publicly
available XENONnT detector simulation
\texttt{fuse}~\cite{henning_schulze_eissing_2026_fuse}. NRE is chosen over
alternative SBI algorithms, such as neural posterior estimation~(NPE), which
directly learns a generative model of the
posterior~\cite{Deistler_2025_SBI-Guide}, because the learned likelihood ratio
is prior-independent and can be more easily compared with conventional
frequentist approaches in the context of exclusion limit setting.

Beyond demonstrating that NRE yields exclusion limits compatible with the
established PLR approach, our central contribution is a systematic study of
astrophysical uncertainty marginalization across a range of halo models~--- from
extreme $\pm5\sigma$ SHM variations to observationally motivated models
including the SHM$^{++}$~\cite{Evans_2018_SHM++} and a Large Magellanic
Cloud~(LMC) contribution~\cite{Cordova_2024_LMC}. A key advantage is
scalability: incorporating additional nuisance parameters requires only
diversifying the training data, leaving the cost of inference unchanged.

The paper is organized as follows. \Cref{sec:sbi} introduces the simulation
pipeline, the data representation, and the NRE algorithm.
\Cref{sec:results_sonly} presents signal-only results together with a discussion
of posterior reliability. \Cref{sec:results_sbg} extends the analysis to the
signal-plus-background case and provides a prior-independent frequentist
cross-check. \Cref{sec:astro} develops the astrophysical uncertainty assessment.
\Cref{sec:discussion} summarizes the results and discusses future extensions.
Appendix~\ref{app:architectures} details the network architectures and
hyperparameters, Appendix~\ref{app:validation} presents an analytical validation
on idealized 1D spectra, and Appendix~\ref{app:diagnostics} defines the
posterior diagnostic tools.

%% =====================================================================
\section{Simulation-Based Inference for Direct Detection}
\label{sec:sbi}
%% =====================================================================

%% ---------------------------------------------------------------------
\subsection{Signal model and simulation pipeline}
\label{subsec:signal}

The differential event rate for WIMP--nucleus scattering per unit detector mass
is~\cite{Lewin_1996_DMDD-Review}
\begin{equation}
    \frac{dR}{dE_R} = \frac{\rho_\chi}{m_\chi m_N}
    \int_{v > v_\mathrm{min}} v\, f(\mathbf{v})\,
    \frac{d\sigma_{\chi N}}{dE_R}(v, E_R)\, d^3v \ ,
    \label{eq:event_rate}
\end{equation}
where $\chi$ denotes the DM particle, $\rho_\chi = 0.3~\mathrm{GeV\,cm^{-3}}$ is
the local DM density, $f(\mathbf{v})$ the DM velocity distribution in the
detector frame, $m_N$ the target nuclear mass, and
$v_\mathrm{min} = \sqrt{m_N E_R / 2\mu_N^2}$ the minimum WIMP velocity required
to produce recoil energy~$E_R$, with $\mu_N$ the WIMP--nucleus reduced mass. We
restrict ourselves to the isoscalar SI
interaction~\cite{Lewin_1996_DMDD-Review}, for which the differential cross
section is
\begin{equation}
    \frac{d\sigma_\mathrm{SI}}{dE_R} =
    \frac{m_N \sigma_\mathrm{SI}^0}{2\mu_N^2 v^2}\, F_\mathrm{SI}^2(q) \ ,
    \qquad \sigma_\mathrm{SI}^0 = \frac{\mu_N^2}{\pi} A^2 c_p^2 \ ,
    \label{eq:si_xsec}
\end{equation}
where $A$ is the nuclear mass number, $F_\mathrm{SI}(q)$ the Helm form
factor~\cite{Lewin_1996_DMDD-Review}, and $q = \sqrt{2 m_N E_R}$ the momentum
transfer.  The coherent $A^2$ enhancement makes xenon~($A\approx131$) a
particularly sensitive target. The effective WIMP--proton coupling $c_p$ has
unit GeV$^{-2}$ and is obtained by integrating out the heavy mediator of the
interaction in the non-relativistic limit.

The two model parameters of interest for our study are
$\boldsymbol{\vartheta} = (\log(m_\chi/\mathrm{GeV}),\, \log(c_p/\mathrm{GeV}^{-2}))$,
where $\log$ denotes $\log_{10}$ throughout the paper. We choose the effective
proton coupling instead of the more commonly used DM-proton cross section
$\sigma_p$, because $c_p$ and $m_\chi$ appear as independent parameters in the
effective Lagrangian and can therefore be assumed as uncorrelated in a Bayesian
analysis, while $\sigma_p$ depends on both $c_p$ and $m_\chi$ as
$\sigma_p = \mu_n^2 c_p^2/\pi$ with $\mu_n$ the WIMP--nucleon reduced mass.
Indeed, while our final results are not prior-dependent, a probability
distribution for the model parameters is needed to generate the training data.
In the present study the two parameters are sampled uniformly over the ranges in
\Cref{tab:param_ranges}. The coupling range corresponds to
$10^{-49}~\mathrm{cm^2} \lesssim \sigma_p \lesssim 10^{-45}~\mathrm{cm^2}$,
which in turn corresponds to $\mathcal{O}(1)$--$\mathcal{O}(100)$ expected
signal events per tonne-year and includes also zero-event spectra, which are
essential for training the classifier to recognize the absence of a signal.

\begin{table}[b]
\centering
\caption{Parameter ranges for spectrum generation.}
\label{tab:param_ranges}
\renewcommand{\arraystretch}{1.1}
\begin{tabular}{ccc}
\toprule
Parameter & Range & Description \\
\midrule
$\log(m_\chi/\mathrm{GeV})$ & $[0,\,3]$ & DM mass \\
$\log(c_p/\mathrm{GeV}^{-2})$ & $[-10.5,\,-8.5]$ & Proton coupling \\
\bottomrule
\end{tabular}
\end{table}

Recoil spectra are computed with WimPyDD~\cite{Jeong_2022_WimPyDD}, which
factorizes the rate into pre-computed nuclear response functions and an
astrophysical halo integral
\begin{equation}
    \eta(v_\mathrm{min}) = \int_{v > v_\mathrm{min}}
    \frac{f(\mathbf{v})}{v}\, d^3v \ ;
    \label{eq:halo_integral}
\end{equation}
the halo model can thus be swapped without recomputing the nuclear physics.
Unless stated otherwise, we adopt the Standard Halo Model~(SHM), in which
$f(\mathbf{v})$ is an isotropic Maxwell--Boltzmann distribution truncated at the
galactic escape velocity and boosted into the detector frame, with a most
probable speed $v_0 = 240~\mathrm{km/s}$ and escape velocity
$v_\mathrm{esc} = 528~\mathrm{km/s}$~\cite{Reid_2014_vesc, GAIA_2018_Gaia};
alternative halo models are introduced in \Cref{sec:astro}.

For each individual simulation, the model parameters are drawn from the prior,
the corresponding differential rate is evaluated on a recoil energy grid, the
expected event count is calculated for a given exposure, the observed event
count is drawn from a Poisson distribution, and individual recoil energies are
then sampled by inverting the cumulative distribution function of the spectrum.
These energies are then passed to
\texttt{fuse}~\cite{henning_schulze_eissing_2026_fuse}, the XENONnT detector
simulation framework, which propagates each recoil through the full detector
response chain to produce the corrected observables~(cS1,~cS2). Because
\texttt{fuse} is built on the same \texttt{strax} data-processing framework used
for real XENONnT data, simulated and real events pass through an identical
reconstruction pipeline.

For the exposure and background, we take the first science run
(SR0)~\cite{XENON_SR0} as our reference, since the more recent science runs
include non-ER background components that our simplified analysis cannot model,
whereas the SR0 background was ER-dominated and is therefore well suited to our
approach. As the goal of this work is a proof of concept rather than an
application to real data, we adopt several simplifications: we set the exposure
to $1.0~\mathrm{tonne\text{-}year}$ (SR0: $1.09$) and model the background as a
single flat ER component in $[0,15]~\mathrm{keV_{ee}}$ with
$\mu_\mathrm{bg} = 150$ expected events. For comparison, SR0 had 152 expected
background events, of which roughly 10\% were non-ER contributions~(accidental
coincidences, surface events, radiogenic neutrons) that are considerably more
difficult to model. The flat shape is motivated by the approximately flat
spectra of the dominant ER sources in XENONnT~(radon daughters, krypton, solar
neutrinos). Background events are generated with \texttt{fuse} and
Poisson-sampled at each simulation call.

\begin{figure}[t]
\centering
\includegraphics[width=0.6\columnwidth]{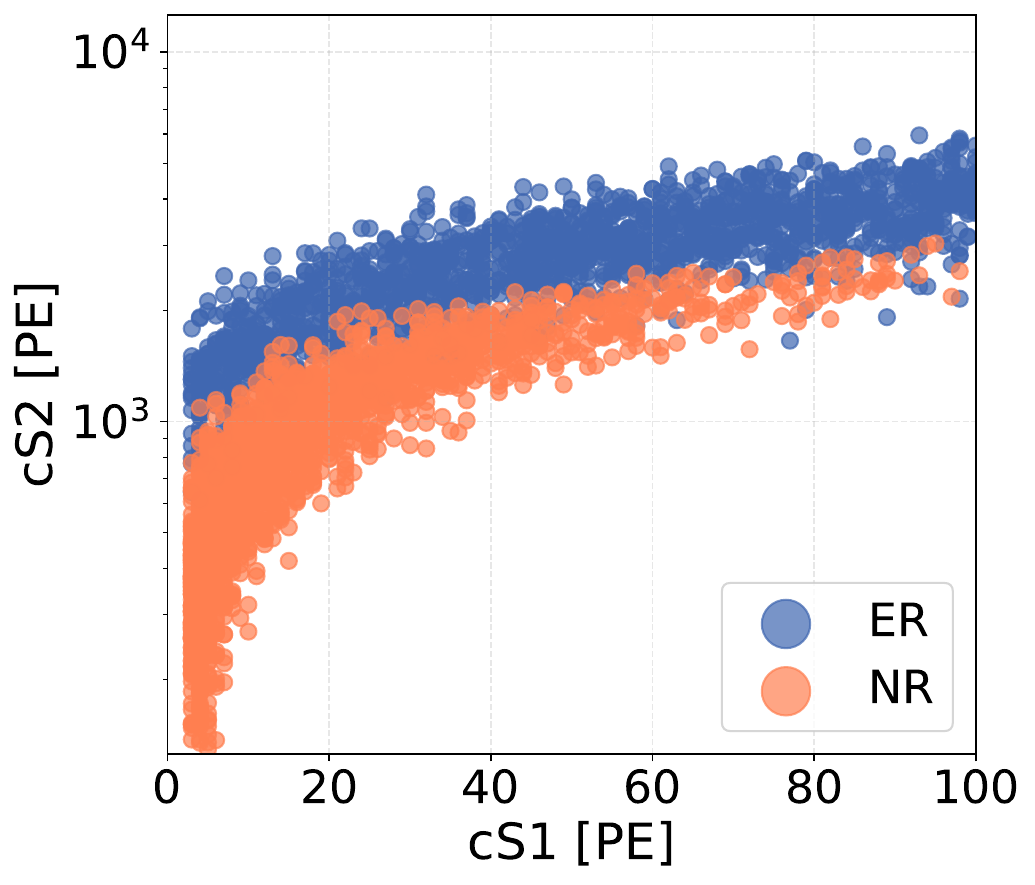}
\caption{Simulated NR signal events~(orange) and flat ER background
  events~(blue) in~(cS1,~cS2) space, drawn from the full range of masses and
  couplings in the training data. The NR and ER bands are well separated due to
  their different charge-to-light ratios, providing the primary discrimination
  variable for the classifier.}
\label{fig:s1s2_scatter}
\end{figure}

%% ---------------------------------------------------------------------
\subsection{Observable representation}
\label{subsec:representation}

The challenge faced by the classifier is illustrated in \Cref{fig:s1s2_scatter}:
NR signal events~(orange) occupy a distinct band in~(cS1,~cS2) space relative to
ER background events~(blue), arising from their different charge-to-light
ratios. This morphological separation provides the primary discrimination
variable for the NRE classifier.

\begin{figure}[t]
\centering
\includegraphics[width=\columnwidth]{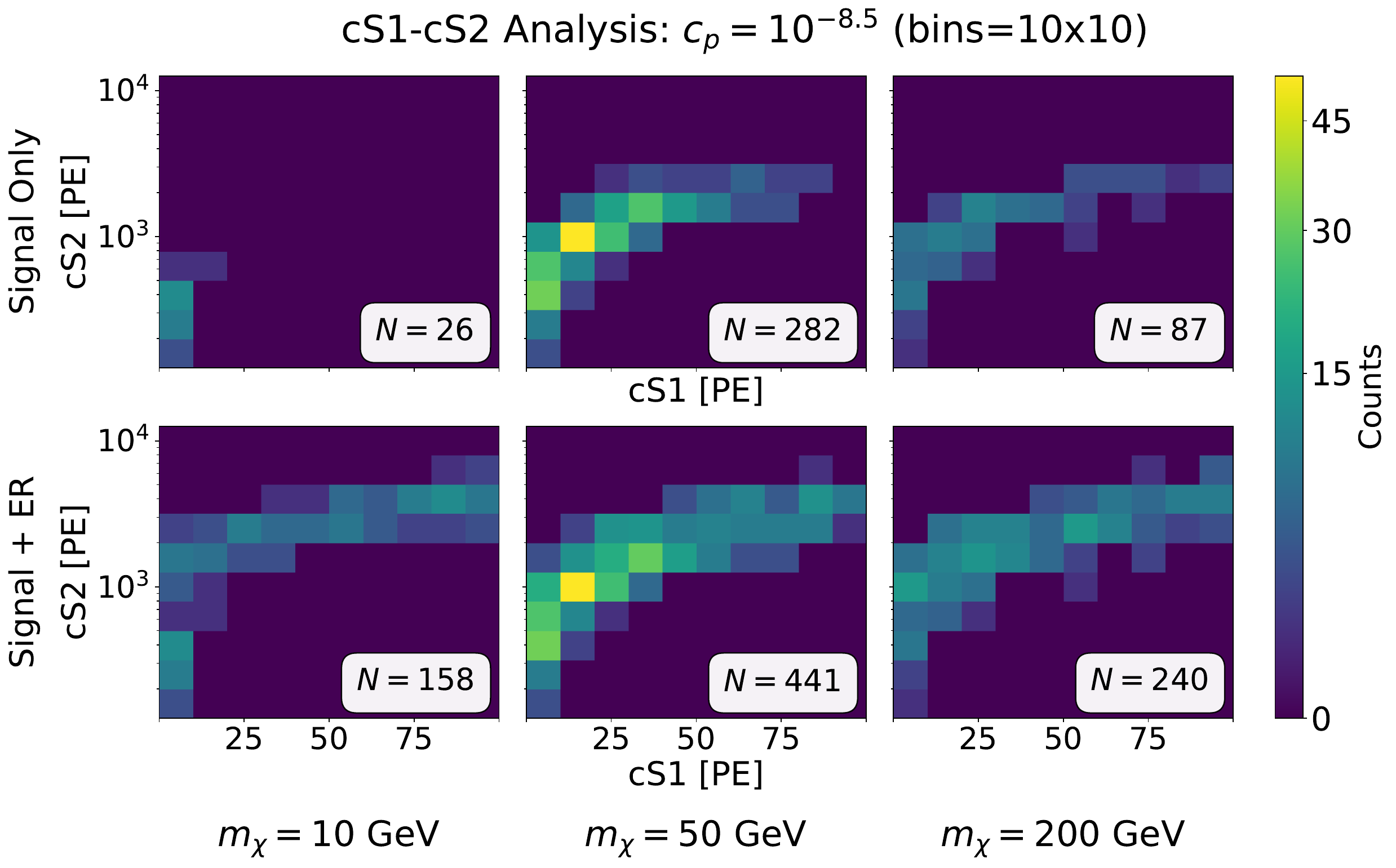}
\caption{Histogrammed~(cS1,~cS2) spectra for $\log c_p = -8.5$ and three DM
  masses~($m_\chi = 10$, 50, $200~\mathrm{GeV}$, columns). Top row: signal only.
  Bottom row: signal plus flat ER background~($\mu_\mathrm{bg}\approx150$).
  Inset boxes indicate total event count~$N$.}
\label{fig:hist_examples_high}

\vspace{5mm}

\includegraphics[width=\columnwidth]{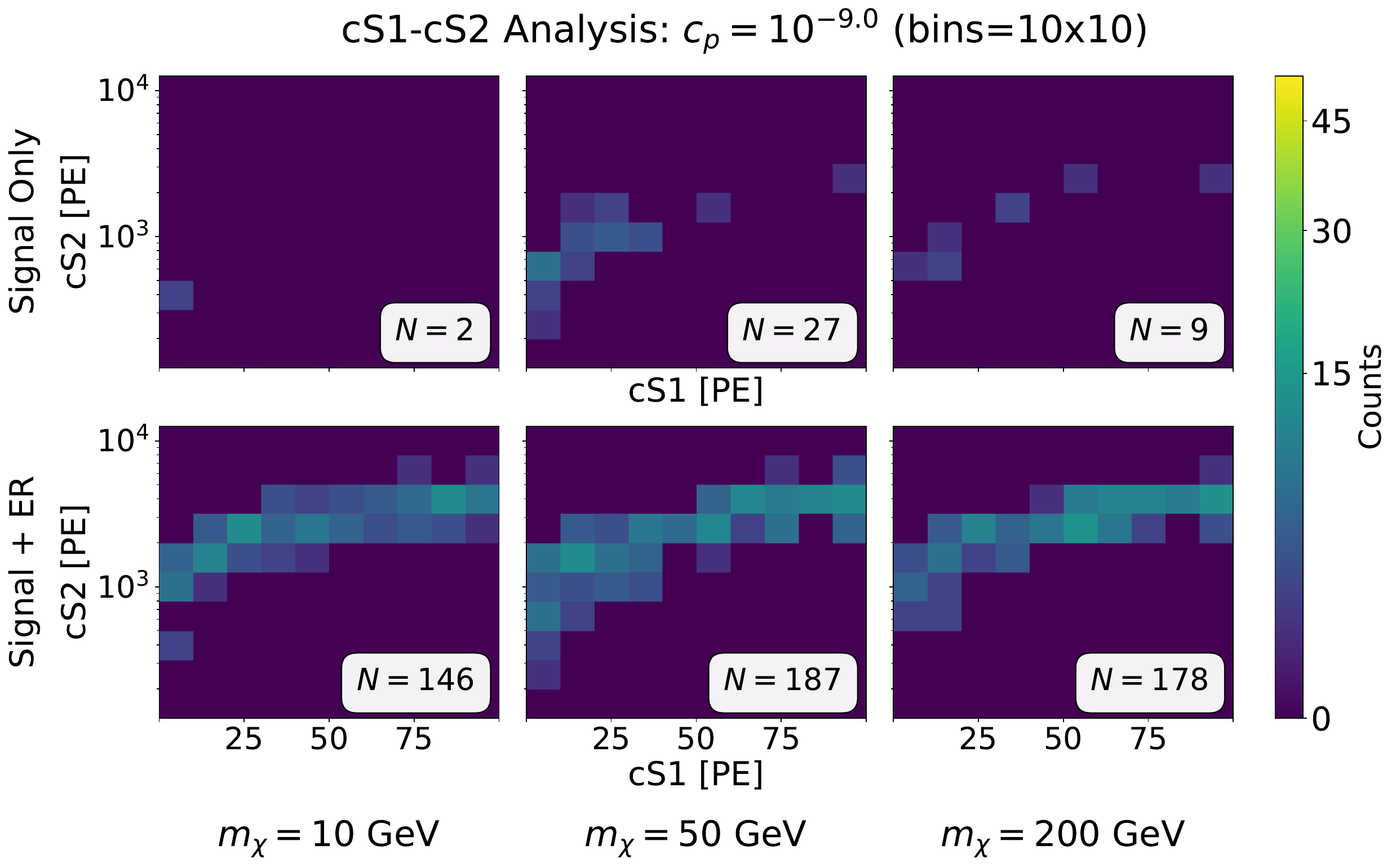}
\caption{Same as \Cref{fig:hist_examples_high} but for $\log c_p = -9.0$. At
  this lower coupling the signal becomes sparse, making discrimination between
  the signal and background rows increasingly subtle.}
\label{fig:hist_examples_low}

\vspace{-5mm}

\end{figure}

The~(cS1,~cS2) event cloud for a given spectrum has a variable number of events.
However, the multilayer perceptron~(MLP) architecture used in this work~(see
\Cref{subsec:nre}) requires a fixed-size input. We therefore bin events into a
two-dimensional histogram~(\texttt{hist\_s1s2}): cS1 binned linearly in
$[0,100]~\mathrm{PE}$ and cS2 logarithmically in
$[10^{2.1}, 10^{4.1}]~\mathrm{PE}$, matching the XENONnT WIMP search region of
interest~\cite{XENON_2026_XENONnt-3.1ty}. This histogram representation is a
natural fixed-size summary of an event cloud, and it retains the full
morphological separation between the NR signal band and the ER background band
while being directly comparable to the observable space of the official XENONnT
analysis. The number of bins per axis and other hyperparameters are discussed in
Appendix~\ref{app:architectures}; we use $n_\mathrm{bins} = 10$ per axis
throughout, so the network input has $10^2 + 2 = 102$ elements after
concatenating with $\boldsymbol{\vartheta}$. All posteriors in this paper are
normalized in log parameter space, consistent with the assumed priors.
Representative histograms are shown in
\Cref{fig:hist_examples_high,fig:hist_examples_low} for strong and weak signals,
respectively. Each training sample for the signal-plus-background classifier
consists of a signal histogram (top rows) combined with an independent
background realization drawn from a large pre-generated pool (bottom rows).

%% ---------------------------------------------------------------------
\subsection{Neural ratio estimation}
\label{subsec:nre}

Our inference framework adopts methods from Bayesian statistics. Given an
observation~$\mathbf{x}$, Bayes' theorem gives the posterior distribution of the
parameters as
\begin{equation}
    p(\boldsymbol{\vartheta}|\mathbf{x}) =
    \frac{p(\mathbf{x}|\boldsymbol{\vartheta})\,p(\boldsymbol{\vartheta})}
         {p(\mathbf{x})} , \ \ 
    p(\mathbf{x}) = \int p(\mathbf{x}|\boldsymbol{\vartheta})\,
                    p(\boldsymbol{\vartheta})\, d\boldsymbol{\vartheta} \ ,
    \label{eq:bayes}
\end{equation}
where $p(\mathbf{x}|\boldsymbol{\vartheta})$ is the likelihood,
$p(\boldsymbol{\vartheta})$ the prior, and the \emph{Bayesian evidence}
$p(\mathbf{x})$ ensures the normalization of the posterior. In our setting,
neither the likelihood nor the evidence has an analytic form, nor can they be
easily inferred numerically from the simulated data.

NRE circumvents this difficulty by learning the likelihood-to-evidence
ratio~(LER)
$r(\mathbf{x}, \boldsymbol{\vartheta}) = p(\mathbf{x}|\boldsymbol{\vartheta}) / p(\mathbf{x})$,
which following \Cref{eq:bayes} is all that is needed to calculate the posterior
from the prior. The LER is estimated with a binary classifier trained to
distinguish \emph{joint}
samples~\mbox{$(\mathbf{x}, \boldsymbol{\vartheta}) \sim p(\mathbf{x}|\boldsymbol{\vartheta})\,p(\boldsymbol{\vartheta})$}~(label
$y=1$) from \emph{marginal}
samples~\mbox{$(\mathbf{x}, \boldsymbol{\vartheta}) \sim p(\mathbf{x})\,p(\boldsymbol{\vartheta})$}~(label
$y=0$)~\cite{Deistler_2025_SBI-Guide}. Marginal samples are constructed by
randomly shuffling parameter labels across the training batch, which breaks the
association between observables and parameters while preserving each marginal
distribution. The Bayes-optimal classifier output~$s$ is directly related to the
likelihood ratio of the two samples:
\begin{equation}
    \frac{s(\mathbf{x}, \boldsymbol{\vartheta})}
         {1 - s(\mathbf{x}, \boldsymbol{\vartheta})}
    = \frac{p(\mathbf{x}|\boldsymbol{\vartheta})\,p(\boldsymbol{\vartheta})}
           {p(\mathbf{x})\,p(\boldsymbol{\vartheta})}
    = r(\mathbf{x}, \boldsymbol{\vartheta}) \; .
    \label{eq:ler}
\end{equation}
Since the prior cancels in the ratio, it can be chosen freely at inference time
without retraining.

For the classifier we use an MLP with SiLU activations and batch normalization,
trained to minimize the binary cross-entropy~(BCE) loss
\begin{equation}
    \mathcal{L}_\mathrm{BCE} = -\frac{1}{N}\sum_{i=1}^{N}
    \bigl[y_i \ln s_i + (1-y_i)\ln(1-s_i)\bigr] \ ,
    \label{eq:bce}
\end{equation}
where $y_i \in \{0,1\}$ is the true label and $s_i$ the predicted score. The MLP
takes the concatenation of the flattened \texttt{hist\_s1s2} histogram and
$\boldsymbol{\vartheta}$ as input and produces a scalar score~$s \in (0,1)$.
Full architecture and hyperparameter details are given in Appendix
\ref{app:architectures}.

Each dataset consists of 300{,}000 simulated spectra. Dataset generation
involves two stages: WimPyDD spectrum generation, taking $\mathcal{O}(300)$ CPU
hours, followed by \texttt{fuse} detector simulation, requiring
$\mathcal{O}(200)$ CPU hours per halo model. All generation workloads are
trivially parallelizable across cores. Classifier training takes
$\mathcal{O}(1\text{--}2)$ CPU hours per model on CPU; GPU acceleration would
provide significant speedup and will be important for more advanced
architectures. Once trained, the classifier evaluates new observations at
negligible cost~(amortized inference).

%% =====================================================================
\section{Signal-Only Results}
\label{sec:results_sonly}
%% =====================================================================

We first train a classifier exclusively on simulated WIMP signal~(cS1,~cS2)
histograms, without any background contribution. This provides the most
optimistic inference scenario and isolates the performance of the method from
the additional difficulty of disentangling signal from background. All results
in this section use the SHM with fixed parameters as the halo model; the impact
of different halo assumptions is studied in \Cref{sec:astro}.

\begin{figure}[t]
\centering
\includegraphics[width=\columnwidth]{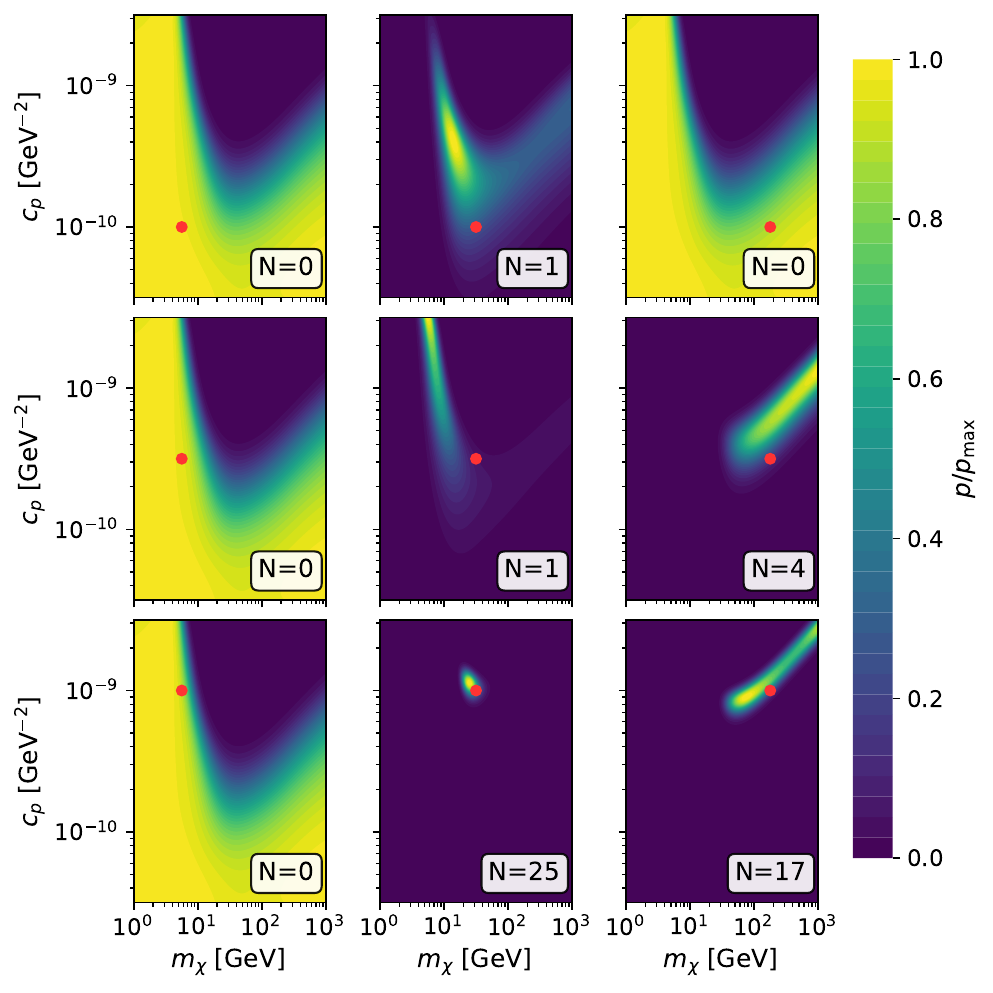}
\caption{NRE posteriors in $(\log m_\chi, \log c_p)$ for the signal-only
  classifier on a $3\times3$ parameter grid, each normalized to its own maximum.
  Red dots mark the true parameter values; inset boxes show the total number of
  detected signal events.}
\label{fig:posteriors_signal}
\end{figure}

\Cref{fig:posteriors_signal} shows NRE posteriors in $(\log m_\chi, \log c_p)$
space on a $3\times3$ grid of parameter points. In each panel, the red dot
indicates the true parameters used for the spectrum generation. We emphasize
that, since the simulations include random fluctuations, it is not expected that
the inferred posterior is centered at the true parameters. As we will show
below, the deviations between the inferred posteriors and the true parameters
agree with the statistical expectation.

Regarding the shape of the posteriors, several qualitative features are
immediately visible in \Cref{fig:posteriors_signal}. At low masses, a sharp
boundary reflects the detection threshold imposed by \texttt{fuse}: recoils with
$E_R \lesssim \mathcal{O}(\mathrm{few\ keV_{NR}})$ are below threshold, so no
events are detected and the posterior cannot constrain small values of~$m_\chi$.
At high masses, the WIMP--nucleus reduced mass saturates~($\mu_N \to m_N$) and
the rate factorizes as $dR/dE_R \propto c_p^2/m_\chi$: different
$(\log m_\chi, \log c_p)$ combinations preserving the total rate become nearly
degenerate, producing a characteristic diagonal ridge in the posterior. The
posterior sharpens with increasing event count, reflecting the improved
statistical discrimination at larger couplings.

%% ---------------------------------------------------------------------
\subsection{Posterior reliability}
\label{subsec:sonly_quality}

Trustworthy inference requires more than a low training loss. The most rigorous
check performed in this study is the comparison to the analytical Poisson
posterior in the idealized 1D case described in Appendix~\ref{app:validation},
which uses true recoil spectra provided by WimPyDD without detector simulation
to determine the event rate per bin deterministically. That validation shows
excellent agreement, confirming that the NRE learns an accurate surrogate. In
the full detector simulation setting, no such analytical reference exists; we
instead rely on coverage tests and distance metrics, described in detail in
Appendix~\ref{app:diagnostics}.

\begin{figure}[t]
\centering
\includegraphics[width=0.9\columnwidth]{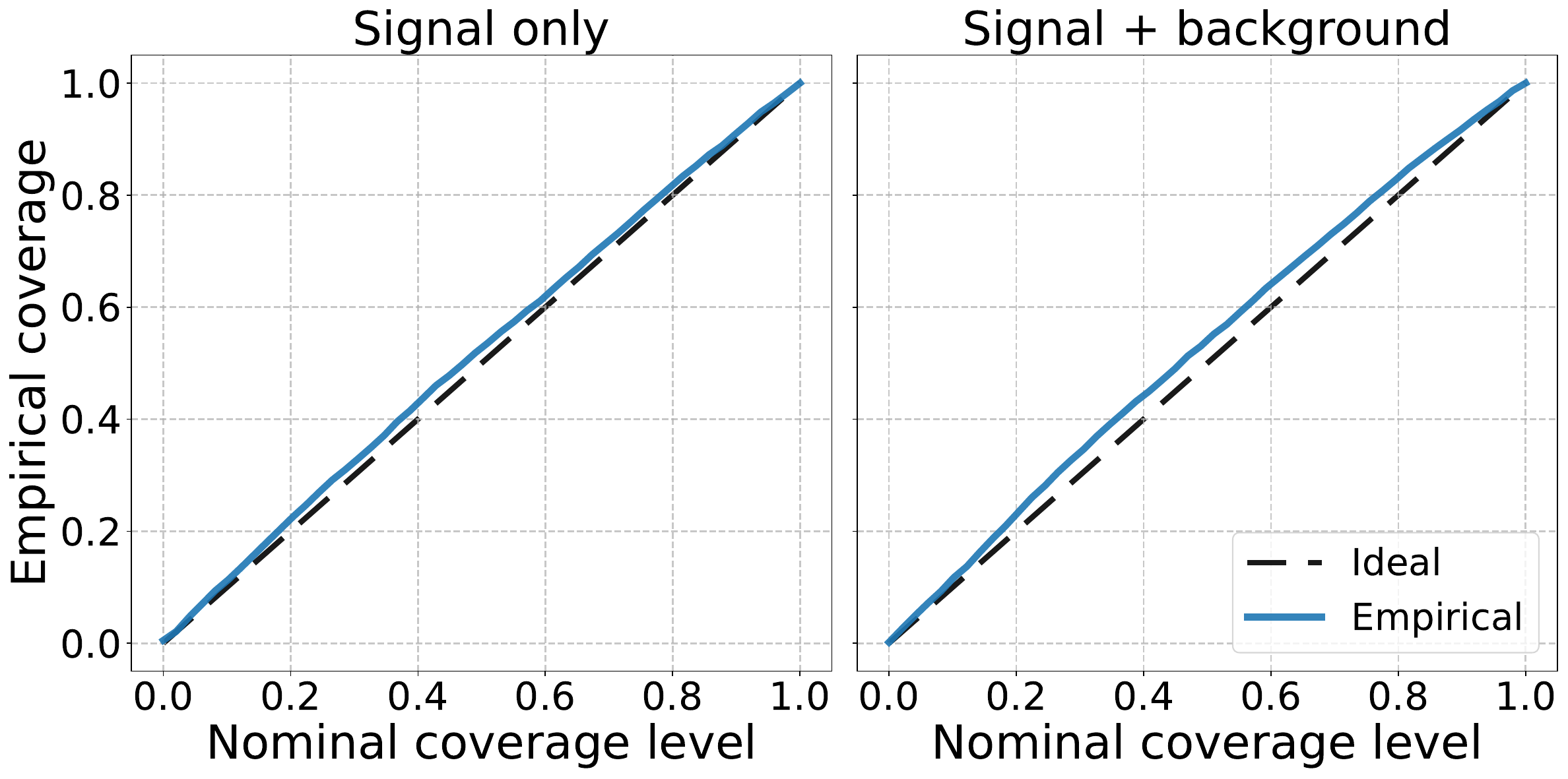}
\caption{Coverage tests for the signal-only~(left) and
  signal-plus-background~(right) classifiers, each evaluated on 5{,}000 test
  spectra. The signal-only curve closely follows the diagonal; the
  signal-plus-background case shows a slight overcoverage due to the broadening
  effect of the ER background.}
\label{fig:coverage_sonly_sbg}
\end{figure}

Coverage tests assess calibration by checking whether true parameter values fall
inside the claimed credible regions with the correct frequency. A
well-calibrated posterior should contain~$\boldsymbol{\vartheta}_\mathrm{true}$
within its $\alpha$-credible highest posterior density~(HPD) region in exactly a
fraction~$\alpha$ of cases; deviations indicate posteriors that are
systematically too broad or too narrow. \Cref{fig:coverage_sonly_sbg} shows
coverage curves for both the signal-only and the signal-plus-background
classifier (discussed below), each evaluated on 5{,}000 test spectra. The
signal-only case follows the diagonal almost perfectly. Including the ER
background introduces a slight overcoverage, since the background adds
stochastic event counts independent of the WIMP parameters and thereby broadens
the posteriors; the resulting inference remains well calibrated and, if
anything, mildly conservative. The Euclidean and Mahalanobis distance metrics
complement coverage tests by assessing maximum a posteriori~(MAP)
point-estimation accuracy; they are discussed in the context of the
astrophysical uncertainty study in Appendix~\ref{app:diagnostics}.

The 90\% HPD exclusion contour from a null~(zero-event) observation is included
in the comparison of \Cref{fig:exclusion_all}, which will be discussed in the
following section. \\

%% =====================================================================
\section{Signal-Plus-Background Results}
\label{sec:results_sbg}
%% =====================================================================

%% ---------------------------------------------------------------------
\subsection{Inference and exclusion procedure}
\label{subsec:sbg_inference}

\begin{figure}[t]
\centering
\includegraphics[width=\columnwidth]{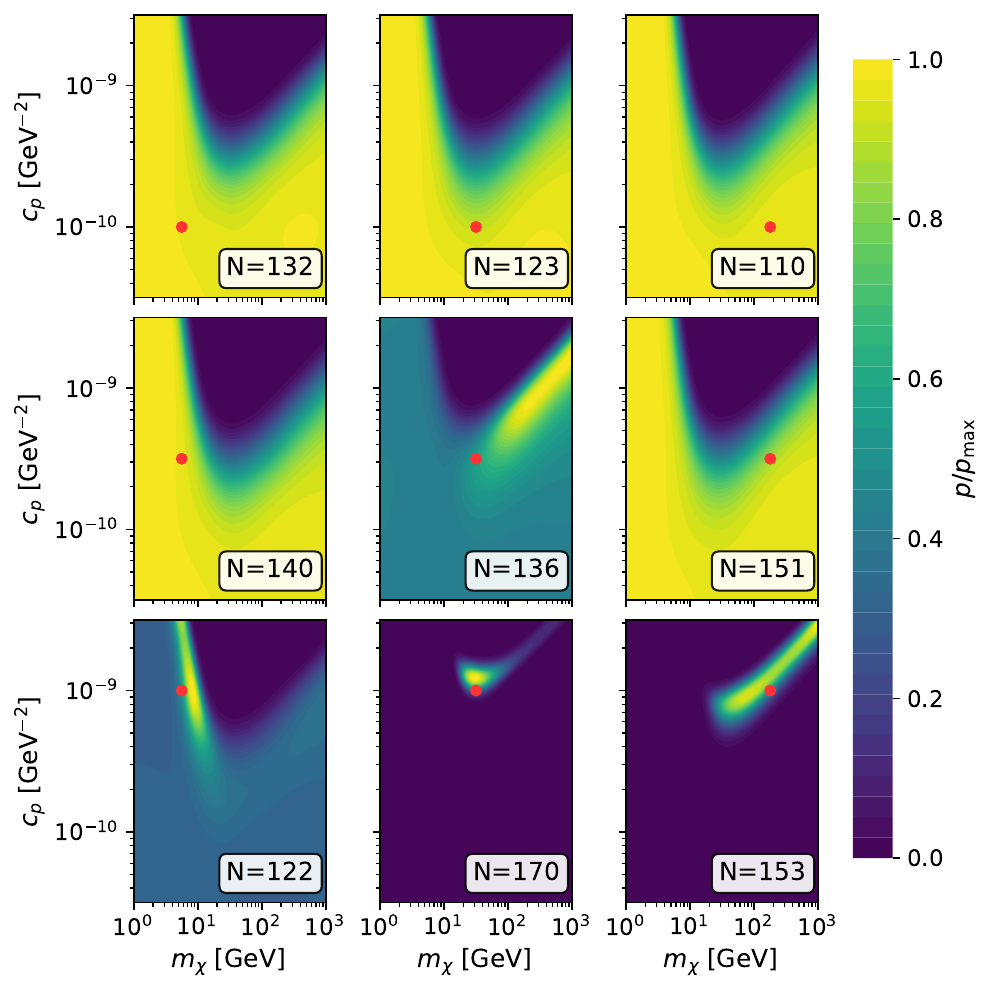}
\caption{Same as \Cref{fig:posteriors_signal} but for the signal-plus-background
  classifier, evaluated on combined WIMP-plus-ER
  spectra~($\mu_\mathrm{bg}\approx150$). Inset boxes show total detected
  events~(signal + background).}
\label{fig:posteriors_sbg}

\vspace{-3mm}

\end{figure}

The signal-plus-background classifier is trained on~(cS1,~cS2) histograms with a
Poisson-sampled flat ER background superimposed on the WIMP signal. The
classifier must learn to disentangle the two contributions using the
morphological ER/NR band separation illustrated in \Cref{fig:s1s2_scatter}. All
results in this section use the SHM as the halo model. \Cref{fig:posteriors_sbg}
shows NRE posteriors on a $3\times3$ parameter grid, evaluated on combined
signal-plus-background spectra. The total event counts in the inset boxes are
dominated by the ER background; posteriors are correspondingly broader than in
the signal-only case, but the diagonal ridge at high masses and the low-mass
threshold boundary remain visible.

A single background realization is insufficient to define a stable exclusion
limit, since a background fluctuation that locally resembles a WIMP signal can
distort the posterior and shift the resulting limit. We therefore average the
posterior over $N = 2000$ independent background realizations to obtain the mean
posterior $\langle p(m_\chi, c_p \,|\, \mathbf{x}) \rangle$, which represents
the expected inference outcome under the background-only
hypothesis.\footnote{Although all posteriors are defined in
$(\log m_\chi, \log c_p)$ space, we drop the logarithms in the notation and
write $p(m_\chi, c_p \,|\, \mathbf{x})$ for brevity.} The 90\% exclusion contour
is then extracted as the boundary of the 90\% HPD region of this averaged
posterior, giving the expected sensitivity of the analysis.

\begin{figure}[t]
\centering
\includegraphics[width=0.85\columnwidth]{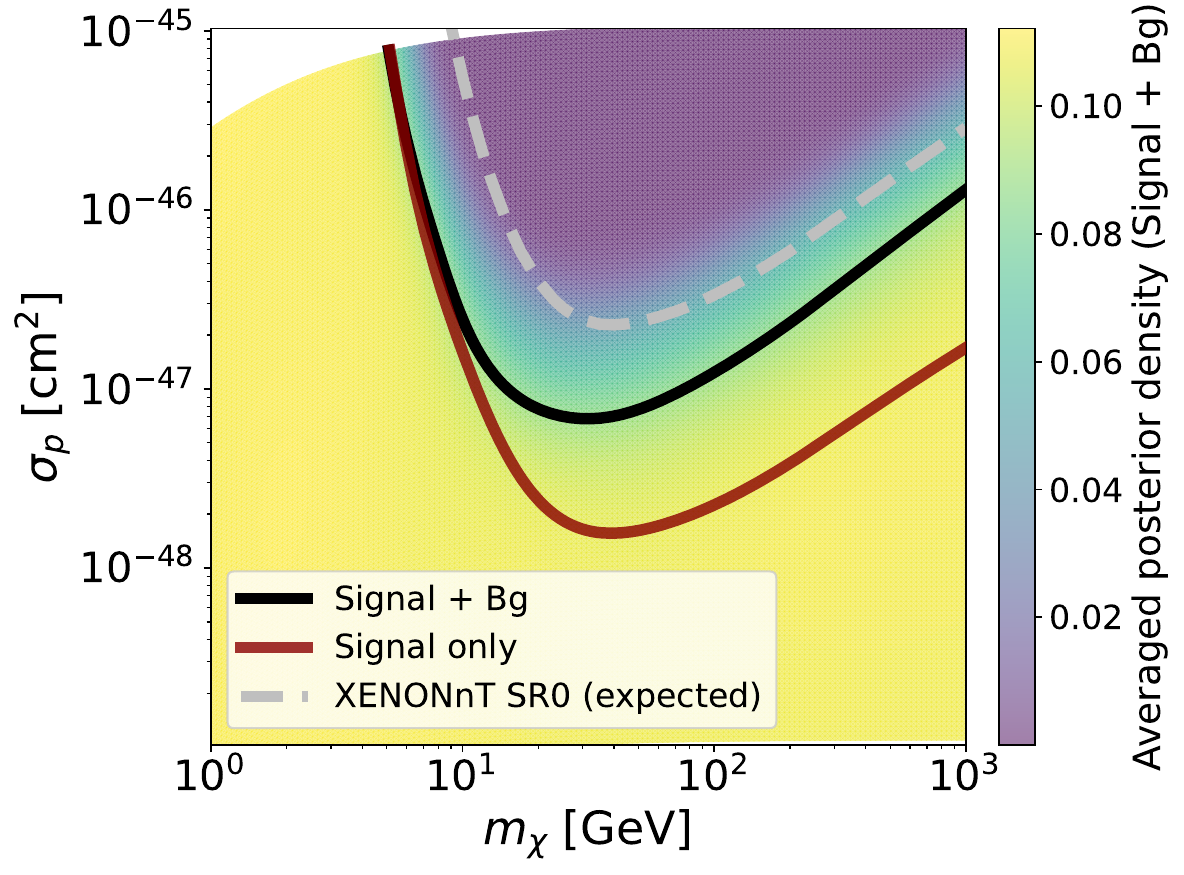}
\caption{90\% CL exclusion contours from the signal-only~(red) and
  signal-plus-background~(black) classifiers. The latter is obtained as the HPD
  boundary of the posterior averaged over $N=2000$ background realizations,
  whose density is underlaid. The XENONnT SR0 median expected
  sensitivity~\cite{XENON_SR0} is shown for reference~(dashed gray).}
\label{fig:exclusion_all}

\vspace{-3mm}

\end{figure}

\Cref{fig:exclusion_all} shows the resulting contours for the signal-only and
signal-plus-background classifiers, with the averaged posterior density
underlaid. The difference between the two curves is physically transparent: the
signal-only contour is the more optimistic of the two, since no background
contamination has to be disentangled from the signal, while the
signal-plus-background contour lies at larger cross sections, reflecting the
additional inference difficulty introduced by the ER background. We also
indicate in \cref{fig:exclusion_all} the expected sensitivity of the first
science run of XENONnT~\cite{XENON_SR0}, which is similar in exposure and
background level to our analysis (see above). We emphasize, however, that these
curves are not directly comparable, since our result is based on a Bayesian
analysis, while the XENONnT expected sensitivity is calculated in the
frequentist framework. In the following, we discuss the impact of the chosen
statistical approach.

%% ---------------------------------------------------------------------
\subsection{Frequentist cross-check and prior sensitivity}
\label{subsec:frequentist}

Even though the NRE classifier learns~$r(\mathbf{x}, \boldsymbol{\vartheta})$,
which carries no prior dependence, the HPD exclusion of \Cref{fig:exclusion_all}
is by construction sensitive to the prior volume, i.e.\ to the chosen parameter
ranges. This is because we implicitly assume the priors (and hence the
posterior) to vanish outside of the shown parameter ranges. If we were to extend
the range of $c_p$ to smaller values, where no signal is predicted, this would
add prior volume exactly in the parameter region where the posterior is largest.
As a result, the 90\% HPD boundary would shift downward in an apparent
sensitivity gain that is an artifact of the prior choice.

\begin{figure}[t]
\centering
\includegraphics[width=0.85\columnwidth]{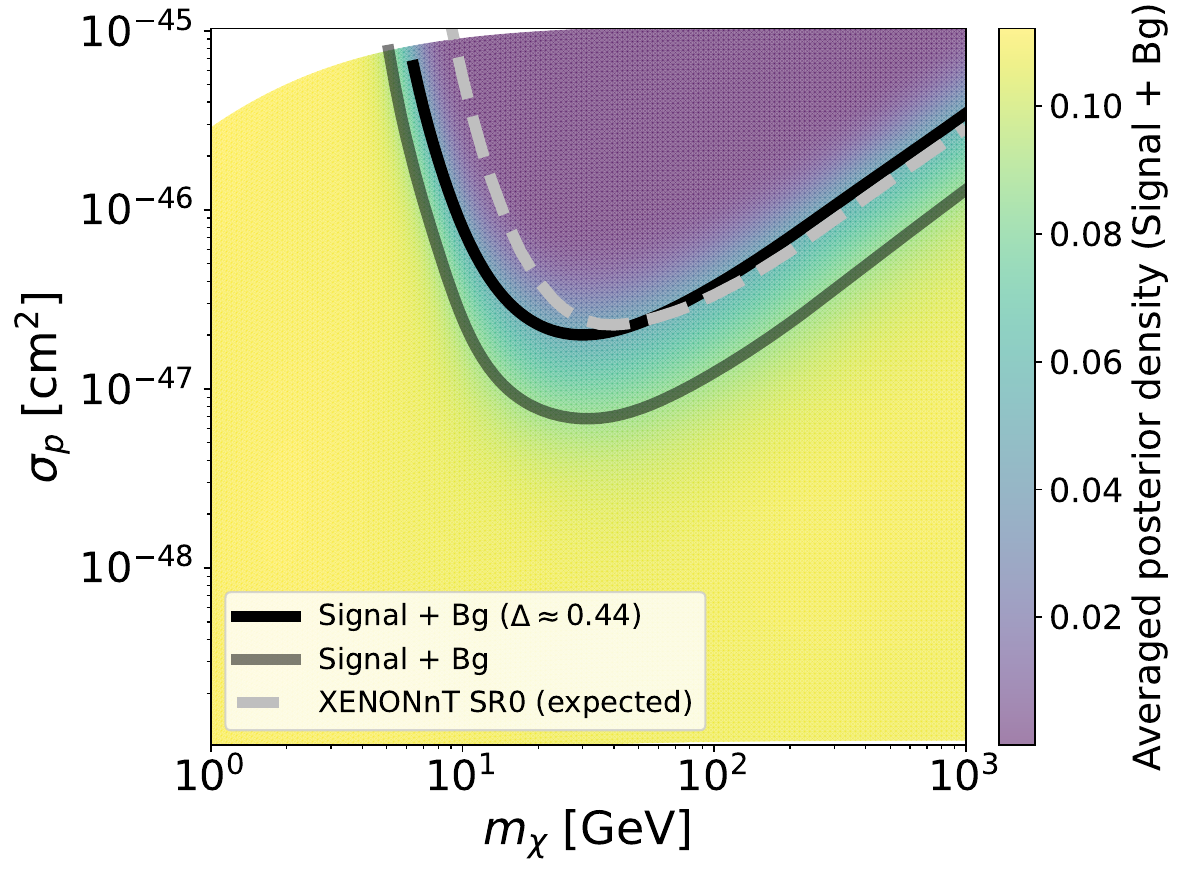}
\caption{Same as \Cref{fig:exclusion_all}, but showing the prior-independent
  90\% CL upper limit from the profiled ratio of \Cref{eq:plr_nre} with
  threshold $\Delta \approx 0.44$~(solid black), restricted to masses above the
  sensitivity threshold. The signal-plus-background HPD contour of
  \Cref{fig:exclusion_all} is shown faded for comparison.}
\label{fig:exclusion_frequentist}
\end{figure}

To obtain a prior-independent limit from the posterior, we can construct a test
statistic analogous to the PLR~\cite{Bringmann:2018jpr}. For each mass $m_\chi$
we define
\begin{equation}
    \lambda(c_p \mid m_\chi) =
    \frac{\langle p(c_p \mid m_\chi, \mathbf{x}) \rangle}
         {\langle p(\hat{c}_p \mid m_\chi, \mathbf{x}) \rangle} \ ,
    \label{eq:plr_nre}
\end{equation}
where $\hat{c}_p$ denotes the value of $c_p$ that maximizes the one-dimensional
posterior. We emphasize that $\lambda$ is independent of the prior ranges and
that the Bayesian evidence cancels in the ratio. The estimator $\hat{c}_p$ is
pushed to zero whenever the data fluctuate below the background-only
expectation. Under the background-only hypothesis, the test statistic
$-2 \ln \lambda$ therefore follows the boundary
mixture~$\tfrac{1}{2}\delta(0) + \tfrac{1}{2}\chi^2_1$~\cite{Conrad_2015_Statistical-Issues},
which leads to a one-sided 90\% threshold of $-2 \ln \lambda \leq 1.64$ or
\begin{equation}
    \lambda \geq \Delta \equiv e^{-1.64/2} \approx 0.44 \ .
    \label{eq:delta}
\end{equation}
For a given DM mass $m_\chi$, the 90\% confidence level upper bound on $c_p$
thus corresponds to the value of $c_p$ at which $\lambda$ is equal to $\Delta$.
We note that this limit can only be obtained above a certain critical mass
$m_\text{min}$. Below this mass, the experiment loses sensitivity and the
averaged posterior becomes flat in~$c_p$, such that $\lambda$ never drops
below~$\Delta$. We determine $m_\text{min}$ as the sensitivity threshold below
which no meaningful limit can be set, and restrict the exclusion curve to higher
masses.

\Cref{fig:exclusion_frequentist} shows the resulting limit alongside the HPD
contour of \Cref{fig:exclusion_all}. The prior-independent limit lies very close
to the XENONnT expected sensitivity based on the PLR method, in particular for
large DM masses. The remaining difference between the NRE result and the XENONnT
expected sensitivity reflects several deliberate simplifications in our
analysis. Most importantly, we have only included a flat ER background, while
the full XENONnT analysis accounts for additional sources~(accidental
coincidences, surface events, radiogenic neutrons) which differ from a DM signal
not only in their (cS1,~cS2) morphology, but also in their spatial distribution,
which is not considered in our analysis.

%% =====================================================================
\section{Astrophysical Uncertainties}
\label{sec:astro}
%% =====================================================================

\begin{table}[b]
\centering
\caption{Halo models used in this analysis. For the SHM, $v_0$ and
  $v_\mathrm{esc}$ are drawn from Gaussian priors at each call; all
  other models use fixed parameters. The LMC velocity distribution is taken
  directly from simulations~\cite{Cordova_2024_LMC}. The extreme models SHMhigh
  and SHMlow represent $\pm5\sigma$ SHM excursions.}
\label{tab:halo_models}
\renewcommand{\arraystretch}{1.1}
\begin{tabular}{lccc}
\toprule
Model & $v_0~[\mathrm{km/s}]$ & $v_\mathrm{esc}~[\mathrm{km/s}]$ & Ref. \\
\midrule
SHM        & $240 \pm 8$ & $528 \pm 24.5$ & \cite{Reid_2014_vesc, GAIA_2018_Gaia} \\
SHM$^{++}$ & 233 & 528 & \cite{Evans_2018_SHM++} \\
LMC        & --- & --- & \cite{Cordova_2024_LMC} \\
SHMhigh    & 280 & 650.5 & --- \\
SHMlow     & 200 & 405.5 & --- \\
\bottomrule
\end{tabular}
\end{table}

The event rate for DM scattering in \Cref{eq:event_rate} depends on the
astrophysical inputs $\rho_\chi$ and $f(\mathbf{v})$. While $\rho_\chi$ enters
only as an overall normalization~\cite{Baxter_2021_DMDD-Recommendations}, the
velocity distribution $f(\mathbf{v})$ enters in a non-trivial way through the
halo integral $\eta(v_\mathrm{min})$ in \Cref{eq:halo_integral}. Uncertainties
in $f(\mathbf{v})$ are most consequential at low DM masses, where the reduced
momentum transfer pushes $v_\mathrm{min}$ into the exponentially suppressed
high-speed tail of $\eta(v_\mathrm{min})$, and different halo models may predict
substantially different rates. In conventional PLR analyses these uncertainties
are usually not included, because doing so would require additional profiling
dimensions and a corresponding increase in toyMC cost. In NRE, by contrast, halo
model nuisance parameters~$\boldsymbol{\eta}$ are handled implicitly and at no
additional inference cost. For each simulation, $\boldsymbol{\eta}$ is drawn
from its prior~$p(\boldsymbol{\eta})$ independently of the particle physics
parameters~$\boldsymbol{\vartheta}$, but the chosen value of $\boldsymbol{\eta}$
is discarded before training. The training data are therefore samples from the
marginal distribution
\begin{equation}
    p(\mathbf{x}, \boldsymbol{\vartheta}) =
    p(\boldsymbol{\vartheta})\int p(\mathbf{x}|\boldsymbol{\vartheta},
    \boldsymbol{\eta})\, p(\boldsymbol{\eta})\, d\boldsymbol{\eta} \ ,
    \label{eq:marginal}
\end{equation}
and the learned surrogate automatically approximates the marginalized
likelihood~$p(\mathbf{x}|\boldsymbol{\vartheta})$, with the halo uncertainty
absorbed into the diversity of the training data. For continuous nuisance
parameters, such as the two parameters~$v_0$ and~$v_\mathrm{esc}$ of the SHM,
$p(\boldsymbol{\eta})$ typically takes the form of a Gaussian distribution that
reflects a priori knowledge, for example from fits to stellar kinematics.
However, it is also possible to consider several discrete halo models, obtained
for example from numerical simulations, in which case $p(\boldsymbol{\eta})$
parametrizes the probability that a specific halo model is selected in the
simulation.

We consider the halo models summarized in \Cref{tab:halo_models}. The SHM serves
as the baseline reference throughout. The SHM$^{++}$~\cite{Evans_2018_SHM++}
adds a radially anisotropic \emph{Gaia Sausage} component reflecting the
kinematic signature of an ancient dwarf galaxy merger. The LMC
model~\cite{Cordova_2024_LMC} accounts for the perturbation of the local DM
distribution by the first pericentric passage of the Large Magellanic Cloud. To
probe the robustness of the inference pipeline under severe misspecification, we
additionally construct two extreme benchmarks as $\pm5\sigma$ excursions in the
SHM nuisance parameters~\mbox{($v_0$, $v_\mathrm{esc}$):} SHMhigh, with a
velocity distribution pushed to higher velocities, and SHMlow, with lower
velocities. We emphasize that these models are not physically motivated but
serve to illustrate the implicit marginalization procedure.

\begin{figure}[t]
\centering
\includegraphics[width=\columnwidth]{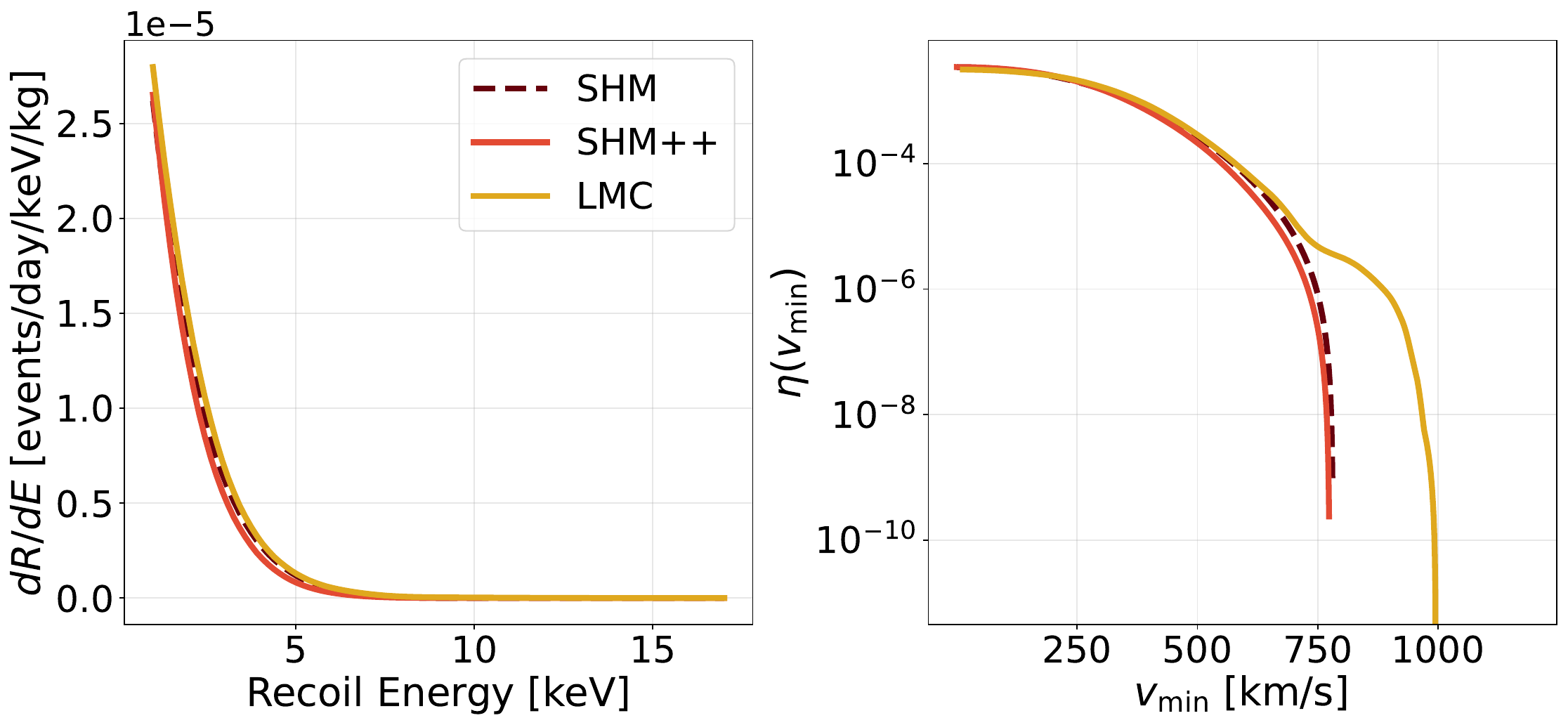}
\caption{Differential event rate for $m_\chi = 10 \, \mathrm{GeV}$ and
  \mbox{$c_p = 10^{-9}~\mathrm{GeV}^{-2}$ (left)} and halo
  integral~$\eta(v_\mathrm{min})$ (right) for SHM, SHM$^{++}$, and LMC. The LMC
  model shows a high-speed tail extending more than $150~\mathrm{km/s}$ beyond
  the SHM-based models.}
\label{fig:halo_comparison}
\end{figure}

\Cref{fig:halo_comparison} illustrates the resulting spectral differences for
the three realistic models. We fix $m_\chi = 10~\mathrm{GeV}$, since for small
DM masses halo model uncertainties have the largest impact on the predicted
rate. The LMC model exhibits a high-speed cutoff shifted by more than
$150~\mathrm{km/s}$ relative to the SHM-based models, arising from DM particles
stripped from the LMC that dominate the high-velocity tail of $f(\mathbf{v})$.
The SHM and SHM$^{++}$ are nearly indistinguishable in the recoil spectrum even
for $m_\chi = 10 \, \mathrm{GeV}$. For larger masses all three models converge
rapidly.

%% ---------------------------------------------------------------------
\subsection{Extreme halo model benchmarks}
\label{subsec:extreme}

Before considering the three realistic halo models, we first consider the two
extreme choices SHMhigh and SHMlow to illustrate the marginalization procedure.
We train three separate classifiers: using only SHMhigh, using only SHMlow, and
using a combined dataset containing both halo models with equal probability. In
all cases we train the classifiers both on signal-only simulations and
signal-plus-background simulations.

\begin{figure}[t]
\centering
\includegraphics[width=\columnwidth]{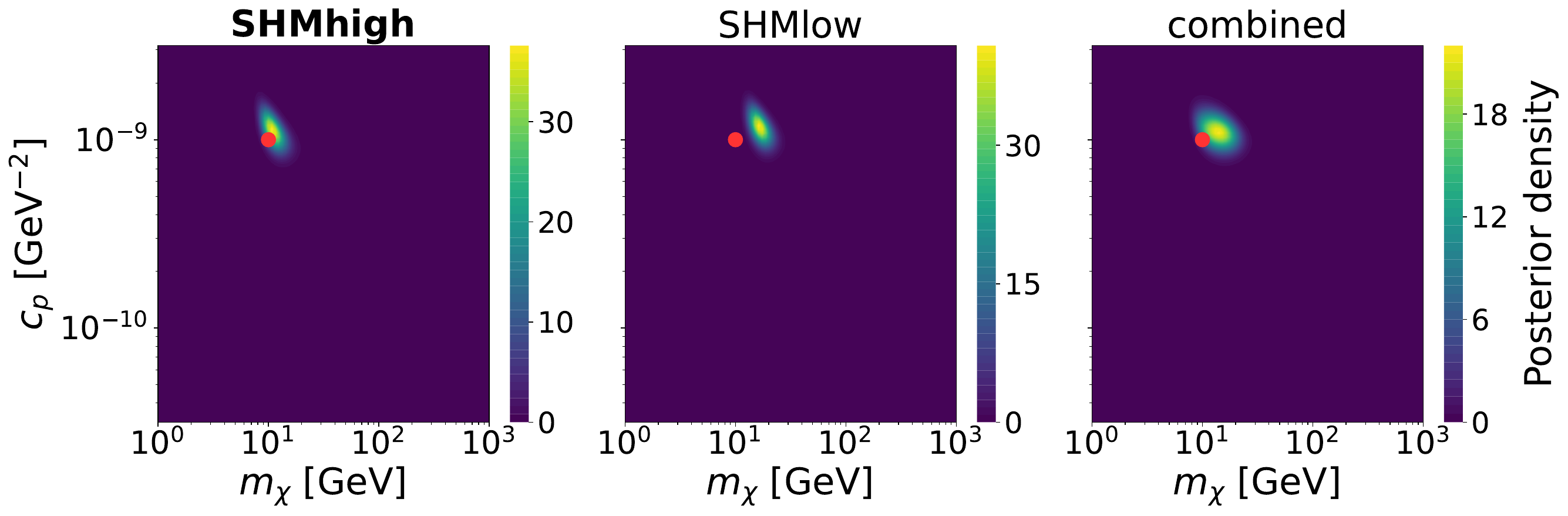}
\caption{Signal-only posteriors for classifiers trained on SHMhigh~(left),
  SHMlow~(middle), and the combined dataset~(right), evaluated on a spectrum
  generated under SHMhigh. The SHMlow posterior is shifted to higher masses; the
  combined posterior broadens to accommodate both interpretations.}
\label{fig:xtreme_posteriors}
\end{figure}

\Cref{fig:xtreme_posteriors} shows a representative example of the posteriors
from the three classifiers evaluated on a signal-only spectrum generated for
SHMhigh. While the classifier trained on SHMhigh yields a posterior that is
compatible with the assumed DM parameters (red dot), the SHMlow classifier
produces a biased posterior that is systematically shifted to higher masses. The
reason is that SHMlow predicts lower halo velocities and hence a softer recoil
spectrum, so observations generated under the SHMhigh distribution are
attributed to a heavier WIMP. The classifier trained on the combined dataset
allows for both possible interpretations and produces a visibly broader
posterior that includes the true model parameters.

\begin{figure}[t]
\centering
\includegraphics[width=0.95\columnwidth]{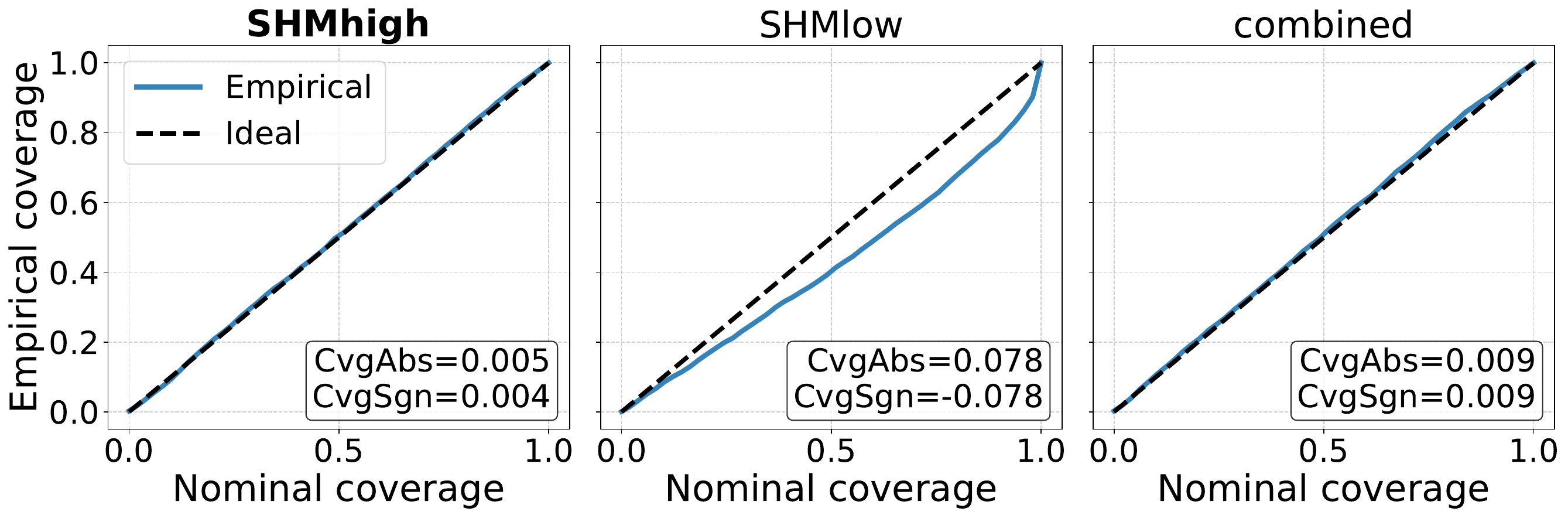}
\vspace{1mm}
\caption{Coverage tests for signal-only classifiers trained on SHMhigh, SHMlow,
  and the combined dataset, evaluated on 5{,}000 test spectra generated under
  SHMhigh.}
\label{fig:sonly_coverage_xtreme}
\end{figure}

These qualitative observations can be made quantitative by considering the
coverage curves shown in \Cref{fig:sonly_coverage_xtreme}, which have been
obtained by evaluating the three classifiers on 5{,}000 test spectra using the
SHMhigh distribution. The correctly specified SHMhigh classifier achieves
excellent calibration, while the misspecified SHMlow classifier shows clear
undercoverage: its posteriors are systematically shifted away from the true
parameters, so the true values fall outside the claimed credible regions more
often than they should. The combined classifier recovers good calibration by
broadening its posteriors. The same qualitative pattern is found using the
signal-plus-background simulations, with the correctly specified and combined
classifiers shifted toward mild overcoverage due to the additional stochastic background
contribution.

\vspace{4mm}
Appendix~\ref{app:diagnostics} provides additional details on how the posteriors
change under halo model misspecification and under the combination of different
halo models. In particular, we quantify in \Cref{tab:xtreme_metrics} the
point-estimation accuracy using Mahalanobis distances. The SHMlow model shows
substantially elevated Mahalanobis distances, reflecting the systematic
posterior shift seen in \Cref{fig:xtreme_posteriors}, while the combined
classifier achieves the lowest Mahalanobis distance despite a larger posterior
volume. These results confirm that implicit marginalization corrects posterior
bias without sacrificing accuracy.

%% ---------------------------------------------------------------------
\subsection{Realistic halo models}
\label{subsec:realistic}

We now repeat the analysis for the three realistic halo models SHM, SHM$^{++}$,
and LMC, with SHM as the reference, and additionally train a combined
classifier, for which all three halo models are mixed with equal probability.

The signal-only posteriors for the three models are visually indistinguishable,
consistent with the nearly identical recoil spectra shown in
\Cref{fig:halo_comparison}. Coverage tests confirm that all three classifiers
are well calibrated, with curves closely following the diagonal. Even the LMC
model, which introduces the largest spectral differences through its shifted
high-speed tail, shows only minimal deviation. In contrast to the extreme cases
discussed above, where the SHMlow showed clear undercoverage and strongly
elevated Mahalanobis distances, all realistic models remain statistically
indistinguishable from the SHM reference in the current sensitivity regime.

\begin{figure}[t]
\centering
\includegraphics[width=0.8\columnwidth]{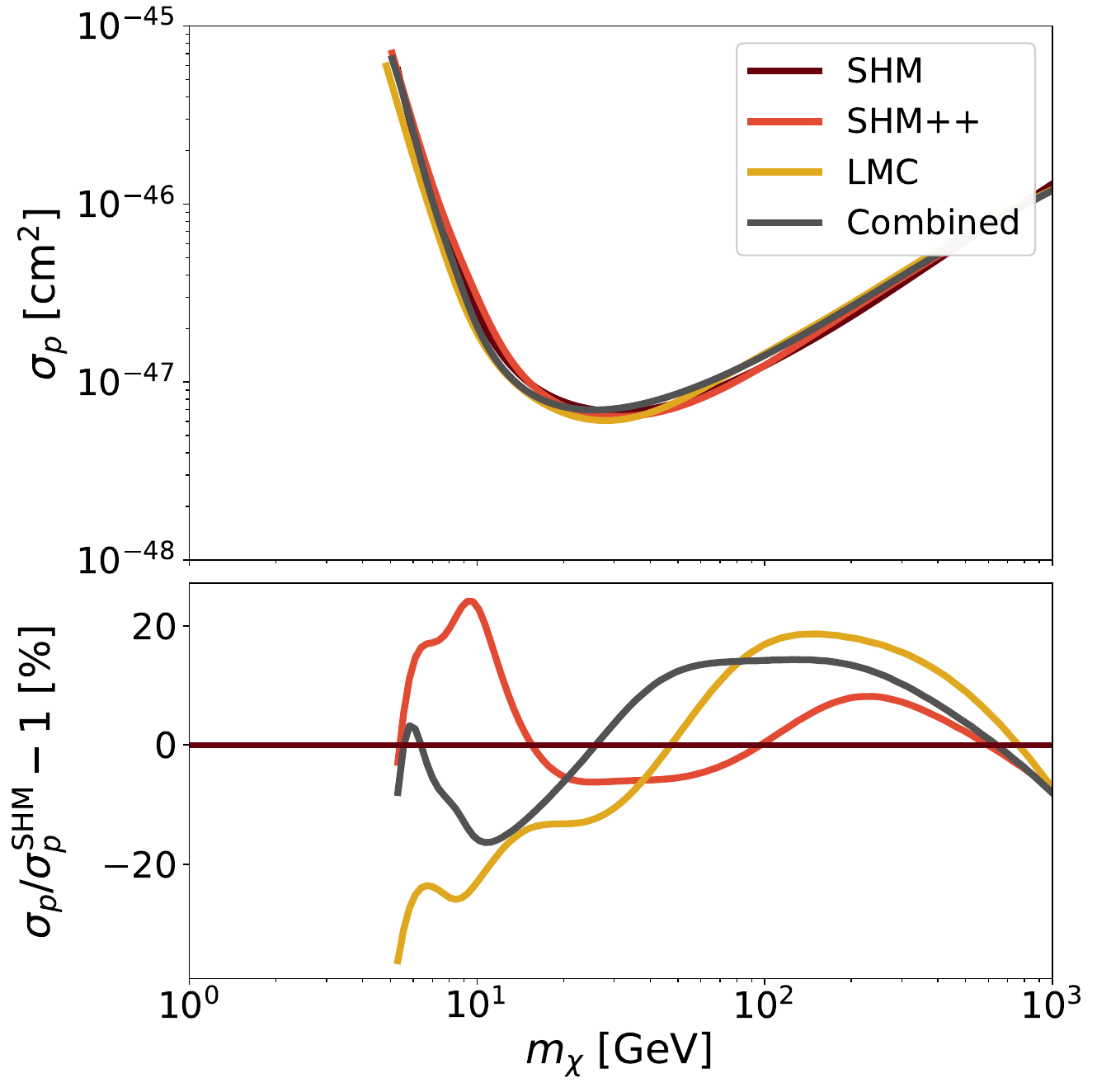}
\caption{Top: 90\% CL median exclusion contours for SHM, SHM$^{++}$, LMC, and
  the combined classifier, obtained from the averaged posterior over $N=2000$
  background realizations. Bottom: relative deviation
  $\sigma_p / \sigma_p^\mathrm{SHM} - 1$ from the SHM reference. Differences are
  most pronounced at low masses, reaching $\pm20\%$.}
\label{fig:real_exclusions}
\end{figure}

\Cref{fig:real_exclusions} shows the exclusion contours from classifiers trained
on signal-plus-background simulations for the realistic models, together with
the relative deviation $\sigma_p / \sigma_p^\mathrm{SHM} - 1$ from the SHM
reference. The contours are very similar across all three models, as expected
from the nearly identical posteriors and coverage results. Differences are most
pronounced at low masses, where distinct predictions for the high-velocity tail
of $f(\mathbf{v})$ shift the contour at the $\pm20\%$ level. The combined
classifier marginalizes over this uncertainty automatically and lies within the
spread of the individual models, without requiring any modification of the
inference step. These results clearly demonstrate that SBI based on a
combination of different halo models constitutes a highly robust and flexible
way to include astrophysical uncertainties in the analysis of direct detection
experiments.

%% =====================================================================
\section{Discussion and Conclusions}
\label{sec:discussion}
%% =====================================================================

In this work we have demonstrated that simulation-based inference is a powerful
tool for the analysis of direct detection experiments. Specifically, we have
applied neural ratio estimation (NRE) using the XENONnT experiment as an example
and provided a proof of concept that the resulting exclusion limits are
qualitatively consistent with those obtained using the conventional profile
likelihood ratio (PLR) approach without the need for an analytic likelihood
function or expensive Monte Carlo simulations.

The signal-plus-background NRE classifier trained on simulated data yields a
90\% CL sensitivity projection that approximates the official XENONnT expected
sensitivity. Most of the deviation can be attributed to inherent differences
between the Bayesian and frequentist approaches and largely disappears in the
frequentist cross-check via the profiled likelihood ratio of \Cref{eq:plr_nre},
which removes prior volume effects. The residual difference reflects our
simplified background model and the fact that no position information is used in
our analysis. We conclude that the NRE procedure recovers the same physical
information as the PLR approach at potentially lower simulation cost.

Indeed, the key advantage of NRE over the PLR approach is its scalability. In
NRE, additional nuisance parameters are incorporated simply by diversifying the
training data, leaving the cost of inference unchanged. This is particularly
valuable as experiments approach the neutrino
fog~\cite{Billard_2014_Neutrino-Fog}, i.e.\ the irreducible background from
coherent elastic neutrino--nucleus scattering~(CE$\nu$NS). In this parameter
region, the correct simultaneous propagation of CE$\nu$NS flux uncertainties,
halo model variations, and instrumental systematics becomes essential.

In the present work we have focused on astrophysical uncertainties from the dark
matter velocity distribution. In NRE, uncertainties in this distribution can be
marginalized over implicitly by combining training datasets from multiple halo
models or from variations in the halo model parameters. Coverage tests on
extreme benchmarks confirm that this procedure corrects undercoverage from halo
misspecification: While a classifier based on an incorrect halo model yields
posteriors that tend to falsely exclude the true model parameters, a classifier
trained on a combination of different models yields well-calibrated results.

We have then applied this approach to realistic halo models based on both
observations and simulations. These models produce nearly indistinguishable
results in the current sensitivity regime, with differences at the
$\mathcal{O}(20\%)$ level in terms of the excluded cross section for low dark
matter masses. The NRE approach removes the need to select one of these halo
models, making it possible to instead include all of them in the analysis
simultaneously.

Our current NRE implementation  has several limitations that point toward a
clear path forward. The coarse $10\times10$ histogram treats the 2D~(cS1,~cS2)
observable as a flat vector, discarding spatial relationships between bins. A
convolutional neural network, as employed in Ref.~\cite{Cerdeno:2024uqt}, would exploit these correlations at much higher
resolution without the quadratic input-dimension scaling that constrains the
multilayer perceptron. Alternatively, a transformer
architecture~\cite{Vaswani_2017_Attention} acting on the unordered set
of~(cS1,~cS2) event pairs could bypass binning entirely.

Before our approach can be applied to real XENONnT data, it will be necessary to
extend the background model to include the full set of background sources with
their rates as nuisance parameters. The background model uncertainties can then
be marginalized over through the same implicit mechanism already employed for
halo uncertainties. However, several of these backgrounds are not uniformly
distributed in the detector volume, such that position information becomes
essential for signal discrimination.

At the same time, it will also be interesting to consider a wider range of
possible particle physics properties of dark matter. Our framework can be easily
extended to other effective operators for dark matter--nucleus scattering beyond
standard spin-independent interactions. In fact, it would also be
straightforward to include several operators simultaneously, for example to
interpret potential excesses in the data (see Ref.~\cite{Cerdeno:2024uqt}). Such excesses may well appear in the
planned XLZD observatory~\cite{XLZD_2025_XLZD}, which will probe deep into the
atmospheric neutrino fog. This is precisely the regime where simulation-based
inference with its implicit marginalization over growing systematic
uncertainties offers its greatest practical advantage.

%% =====================================================================
\begin{acknowledgments}

We thank Klaus Eitel, Ulrich Husemann, Stephen Jiggins, Judith Katzy, Belina von
Krosigk, Teresa Marrod\'an, Kylian Schmidt and Kathrin Valerius for valuable
discussions, and Sebastian Vetter for extensive help with \texttt{fuse}, a
software framework developed by the XENON collaboration.

\end{acknowledgments}

%% =====================================================================
\appendix

\section{Neural Network Architectures and Training}
\label{app:architectures}

\Cref{tab:hpo_all} collects the hyperparameter configurations for the three
primary classifiers: the signal-only and signal-plus-background classifiers
trained on (cS1, cS2) data and discussed in the main text, and the idealized
classifier trained on WimPyDD spectra and discussed in
Appendix~\ref{app:validation}. Hyperparameters are optimized with
Optuna~\cite{Akiba_2019_Optuna} using the Tree-structured Parzen Estimator~(TPE)
sampler and a median pruner, running for up to $\sim200$ trials per classifier.
The TPE sampler builds a probabilistic model of the objective function to
propose promising configurations, while the median pruner terminates unpromising
trials early based on intermediate validation losses. All models use SiLU
activations, batch normalization after each hidden layer, AdamW
optimization~\cite{Loshchilov_2019_AdamW} with an exponential learning rate
decay, and early stopping with a patience of 20 epochs. The training split is
80\% training, 15\% validation, 5\% test.

The number of histogram bins~$n_\mathrm{bins}$ for the~(cS1,~cS2) classifiers
was not included in the formal hyperparameter optimization. Since the MLP treats
the histogram as a flat vector, spatial correlations between neighboring bins
are not exploited, and finer binning primarily increases the input dimension
quadratically without a commensurate gain in discriminating power. We therefore
fix $n_\mathrm{bins} = 10$ throughout; CNN-based architectures, which can
exploit local spatial structure, would be better suited to benefit from higher
resolutions. The progression of best validation losses~--- $0.376$~(idealized
1D) $\to 0.411$~(full detector, signal only) $\to 0.585$~(signal plus
background)~--- quantifies the incremental inference cost of each modeling step.

\begin{table}[t]
\centering
\caption{Hyperparameter specifications for the three primary NRE classifiers.}
\label{tab:hpo_all}
\renewcommand{\arraystretch}{1.1}
\begin{tabular}{lccc}
\toprule
 & WimPyDD & \multicolumn{2}{c}{(cS1,~cS2)} \\
\cmidrule(lr){2-2} \cmidrule(lr){3-4}
Parameter & idealized & sig.\ only & sig.+bg \\
\midrule
Input dim.      & 13 & 102 & 102 \\
Histogram       & --- & $10\times10$ & $10\times10$ \\
Hidden layers   & 8 & 5 & 6 \\
Units/layer     & 512 & 128 & 64 \\
Dropout         & 0.05 & 0.10 & 0.20 \\
LR              & $3\!\times\!10^{-3}$ & $3\!\times\!10^{-3}$ & $3\!\times\!10^{-3}$ \\
Weight decay    & $10^{-6}$ & $10^{-3}$ & 0.0 \\
Batch size      & 2048 & 1024 & 1024 \\
Optuna trials   & 150 & 176 & 232 \\
Best val.\ loss & 0.376 & 0.411 & 0.585 \\
\bottomrule
\end{tabular}
\end{table}

\section{Analytical Validation on Idealized 1D Spectra}
\label{app:validation}

Before applying NRE to data from full detector simulations, we validate the
method on one-dimensional WimPyDD nuclear recoil spectra without detector
effects. Bypassing \texttt{fuse} and working directly in the 1D recoil energy
space has an important practical advantage: because WimPyDD provides the
expected event rate per bin~$\lambda_b(\boldsymbol{\vartheta})$
deterministically, the Poisson likelihood for a binned observation
$\mathbf{n} = (n_1,\ldots,n_B)$ is analytically tractable,
\begin{equation}
  \ln \mathcal{L}(\boldsymbol{\vartheta})
  = \sum_{b=1}^{B}
    \bigl[ n_b \ln \lambda_b(\boldsymbol{\vartheta})
           - \lambda_b(\boldsymbol{\vartheta}) \bigr] \ .
  \label{eq:poisson_binned}
\end{equation}
Combined with a flat prior, this yields an analytical reference posterior
against which the NRE output can be directly compared. We quantify the agreement
using the Jensen--Shannon divergence~(JSD), defined as
\begin{align}
  \mathrm{JSD}(P \| Q) &=
  \tfrac{1}{2}\, D_\mathrm{KL}(P \| M)
  + \tfrac{1}{2}\, D_\mathrm{KL}(Q \| M) \ ,
  \notag \\
  M &= \tfrac{1}{2}(P+Q) \ ,
  \label{eq:jsd}
\end{align}
where $D_\mathrm{KL}(P\|Q) = \int P\log_2(P/Q)$ is the Kullback--Leibler
divergence. The JSD is symmetric, bounded in $[0,1]$, with $\mathrm{JSD}=0$ for
identical distributions.

For this idealized case we summarize each recoil energy list by a compact,
fixed-size feature vector: the total event count~$N_\mathrm{tot}$ augmented by
the ten highest recoil
energies,~\mbox{$\mathbf{x} = (N_\mathrm{tot}, E_{(1)}, \ldots, E_{(10)}) \in \mathbb{R}^{11}$}.
Spectra with fewer than ten detected events are zero-padded to the fixed length.
$N_\mathrm{tot}$ constrains~$c_p$ through the overall rate $\propto c_p^2$,
while the highest recoil energies carry the spectral shape information most
sensitive to~$m_\chi$. Using only $N_\mathrm{tot}$ performs substantially worse,
because the mass--coupling degeneracy~($dR/dE_R \propto c_p^2/m_\chi$ at high
masses) cannot be resolved without spectral shape information.

\begin{figure}[t]
\centering
\includegraphics[width=\columnwidth]{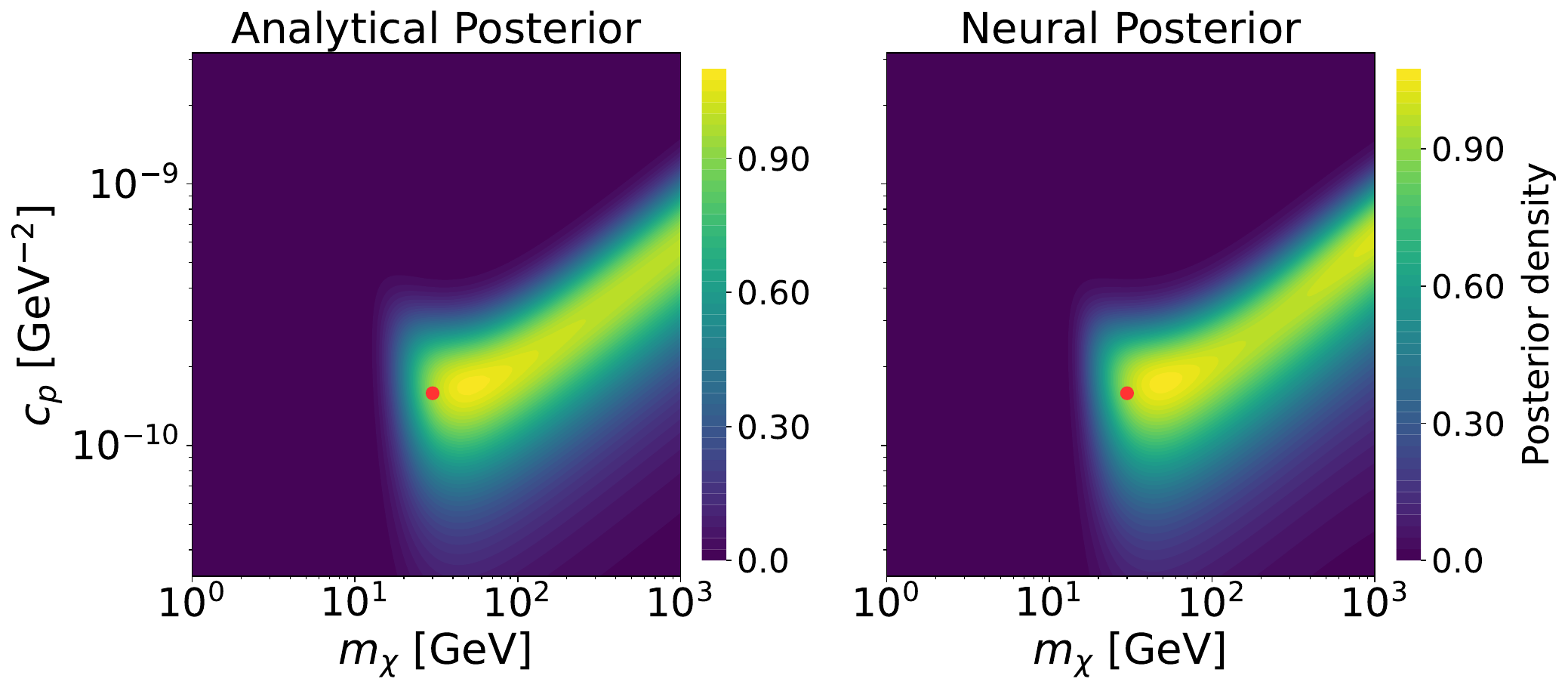}
\caption{NRE posterior~(left) and analytical Poisson posterior~(right) for a
  representative 1D WimPyDD spectrum. The two posteriors agree closely in shape,
  location, and width.}
\label{fig:val_comparison}
\end{figure}

We find that the idealized classifier recovers the analytical posterior with
very high accuracy. Typical values of the JSD are $\mathcal{O}(0.01)$ across the
entire parameter space. Coverage tests furthermore confirm excellent
calibration. \Cref{fig:val_comparison} shows a representative comparison between
NRE and analytical posteriors. We conclude that the classifier is able to
reconstruct the DM properties with negligible loss of accuracy compared to the
conventional likelihood-based approach.

\section{Posterior Diagnostics: Distance Metrics}
\label{app:diagnostics}

Coverage tests, introduced in \Cref{subsec:sonly_quality}, assess calibration
but are not sufficient on their own: a posterior can be well-calibrated yet
uninformatively broad. We therefore complement them with distance metrics that
assess point-estimation accuracy. We use the Euclidean distance between the MAP
estimate and $\boldsymbol{\vartheta}_\mathrm{true}$ in
$(\log(m_\chi/\mathrm{GeV}),\, \log(c_p/\mathrm{GeV}^{-2}))$ space, and the
Mahalanobis distance
\begin{equation}
  d_M = \sqrt{(\hat{\boldsymbol{\vartheta}} -
  \boldsymbol{\vartheta}_\mathrm{true})^\top \Sigma^{-1}
  (\hat{\boldsymbol{\vartheta}} - \boldsymbol{\vartheta}_\mathrm{true})} \ ,
  \label{eq:mahalanobis}
\end{equation}
which normalizes the displacement by the posterior covariance~$\Sigma$, so that
a large MAP error is penalized less when the posterior is correspondingly broad.
For a perfectly calibrated Gaussian posterior~$d_M = 1$ on average. We
additionally track the posterior volume~$V$ relative to a reference model;
classifiers trained on combined datasets are expected to yield broader
posteriors~(larger~$V$) as a natural consequence of marginalizing over a more
complex training distribution. Together, $d_M$ and $V$ distinguish a
broad-but-correct posterior~($d_M \approx 1$, large~$V$) from a
narrow-but-shifted one~($d_M \gg 1$, small~$V$).

\Cref{tab:xtreme_metrics} lists these quantities for the extreme halo model
classifiers discussed in \Cref{subsec:extreme}. These numbers confirm the
conclusion that the misspecified model leads to a shift in the posterior
(increased distance metrics), while the combined model leads to well-calibrated
but broader posteriors (increased posterior volume).

\begin{table}[h]
\centering
%\vspace{5mm}
\caption{Mean distance metrics and posterior volume for signal-only classifiers
  evaluated on 5{,}000 test spectra generated under SHMhigh. $V/V_\mathrm{ref}$
  denotes the posterior volume relative to the SHMhigh classifier.}
\label{tab:xtreme_metrics}
\renewcommand{\arraystretch}{1.1}
\begin{tabular}{lccc}
\toprule
Model & Euclidean & Mahalanobis & $V/V_\mathrm{ref}$ \\
\midrule
SHMhigh    & 0.09 & 1.41 & 1.00 \\
SHMlow     & 0.12 & 1.98 & 0.91 \\
Combined   & 0.10 & 1.45 & 1.08 \\
\bottomrule
\end{tabular}
\end{table}

\FloatBarrier

%% =====================================================================
\bibliography{refs}

\end{document}